%% file: main.tex
\documentclass[twocolumn,aps,prb,showpacs,preprintnumbers,amsmath,amssymb]{revtex4-1}
\usepackage{subfiles}
\usepackage{natbib}
\usepackage{ifthen}
\usepackage{color}
\usepackage{graphicx}
\usepackage{epstopdf}
\usepackage{graphicx}
\usepackage{epsfig}
\usepackage{ulem}
\usepackage{latexsym}
\usepackage{amsmath}
\usepackage{amssymb}
\usepackage{bm}
\usepackage{wasysym}
\usepackage{gensymb}
\usepackage{overpic}
\usepackage{latexsym}
\usepackage{amsmath}
\usepackage{amssymb}
\usepackage{bm}
\usepackage{wasysym}
\begin{document}
\subfile{self_trapped_QED.tex}

\subfile{self_trapped_suppl.tex}
\end{document}

%% file: self_trapped_QED.tex
\title{Engineering indefinitely long-lived localization in cavity-QED arrays}

\author{Amit  Dey}
\address{International Centre for Theoretical Sciences, Tata Institute of Fundamental Research, Bengaluru -- 560089, India}

\author{Manas Kulkarni}
\address{International Centre for Theoretical Sciences, Tata Institute of Fundamental Research, Bengaluru -- 560089, India}

\date{\today}
\begin{abstract}
   By exploiting the nonlinear nature of the Jaynes Cumming's interaction, one can get photon population trapping in cavity-QED arrays. However, the unavoidable dissipative effects in a realistic system would  destroy the self-trapped state by continuous photon leakage. 
 To circumvent this issue, we show that a careful engineering of drive, dissipation and Hamiltonian results in achieving indefinitely sustained self-trapping. We show that the intricate interplay between drive, dissipation, and light-matter interaction results in requiring an optimal window of drive strengths in order to achieve such non-trivial steady states. We treat the two-cavity and four-cavity cases using exact open quantum many-body calculations. Additionally, in the semiclassical limit we scale up the system to a long 1-D chain and demonstrate localization de-localization transition in a driven-dissipative system. Although, our analysis is performed keeping cavity-QED systems in mind, our work is applicable to other driven-dissipative systems where nonlinearity plays a defining role. 

\end{abstract}
\maketitle

\textit{Introduction:}
Intricate interplay between the nonlinear interactions and kinetic hopping delivers fascinating physics of Josephson oscillation and macroscopic quantum self-trapping, which are already realized in bosonic Josephson junction (BJJ) consisting of
cold-atomic Bose-Einstein condensates (BEC) \cite{milburn,smerzi,levy,albeiz}. The novel phenomenon of self-trapping is a consequence of strong nonlinear on-site interaction 
that dominates over the particle tunnelling. Despite short-lived polariton states and weak photon-photon interaction \cite{mabuchi,girvin,natphysrev}, nonlinear Josephson oscillation as well as macroscopic self-trapping have been achieved in 
photonic systems (e.g., cavity-QEDs) characterized by light-matter interactions \cite{shelykh,abbarchi,coto,tureci_prb,raftery}. Photonic Josephson junction (PJJ) has come up as a novel platform for quantum many body simulation.  
Remarkable development in fabrication techniques, precision control of individual components, efficient tunability of parameters, and scalability make cavity-QED systems an important platform for various fields ranging from nonlinear optics to 
quantum communication.

Extended version of PJJ can be realized in a lattice of cavities with nearest-neighbour tunnelling. Realizations of large bosonic systems, such as Bose-Hubbard 
lattice \cite{greiner,bloch,amit} of ultra-cold atoms and strong light-matter coupling \cite{kimble} have motivated exploration of strongly correlated phases in cavity-QED arrays.     
Scaled-up cavity arrays with various coordination numbers can offer exotic collective many body phenomena often missed in few body physics (for e.g., Mott-Insulator).  Large scale systems offers a plethora of possibilities and surprises.  
In Jaynes-Cummings cavity arrays, strong correlations of photons and their collective behavior have been well investigated. Furthermore, in these setups, Mott insulator to superfluid phase transition is investigated \cite{greentree,hartmann,rossini,littlewood}. 
Addressing dissipation engineering in a cavity network, a purely dissipation induced phase transition from superfluid-like state to Mott-insulator-like state was predicted beyond a critical cavity decay \cite{coto,brown}. Additionally, 
drawing correspondence to Bose-Hubbard dynamics, driven-dissipative cavity arrays are presented as efficient quantum simulators \cite{mendoza,lieb} implementable with present technology. 
Moreover, a dissipative phase transition has been observed experimentally in a 1-D lattice of 72 microwave cavities each coupled to superconducting qubit \cite{mattias}. 
Thus substantial progress has been made regarding scalable cavity-QED based architectures \cite{sebastian}. Given these experimental and theoretical advances in large scale systems, exploring the possibility of non-trivial non-equilibrium steady states is an interesting avenue of research.

\begin{figure}[h]
  \centering
   \includegraphics[width=3.4in]{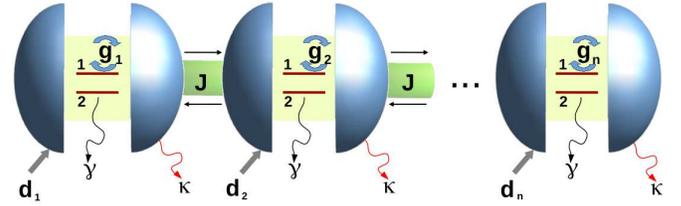}
  \caption{Schematic depiction of 1-D lattice of identical cavity-QEDs each containing a two-level system with levels marked as 1 and 2. Each two-level system is coupled to the cavity photons with a strength $g_i$. Additionally, J,  $\kappa$, $\gamma$ represent inter-cavity tunneling, photon decay, and spontaneous decay, respectively. $d_i$ is the coherent photon microwave drive applied at the i-th cavity.}
  \label{figsch}
\end{figure}

Dissipation-induced delocalization-localization transition of photons was theoretically predicted \cite{tureci_prb} and experimentally 
verified \cite{raftery}. The critical atom-photon interaction for such a transition is large and can be well achieved in a cavity consisting of superconducting qubit coupled to  
transmission line resonators \cite{wallraff,neereja}. Although self-trapping is achieved in a coupled cavity dimer, continuous unavoidable photon leakage and spontaneous decay 
of the qubit limit the longevity of self-trapped state in a realistic system \cite{tureci_prb}. Therefore, it is imperative that competition of dissipation, drive, 
and interaction strength be explored in detail and requirements for robust localization are sought. In fact, the idea of striking a delicate balance between drive and dissipation has been very successful in preparing target states \cite{sg}, achieving indefinitely long-lived entanglement between qubits \cite{mk1a,mk1,mk2,aron1} 
and persistent chiral currents \cite{mk3}. 

Here, we investigate a strong-coupling regime where both drive and dissipation are present and hope to create indefinitely long-lived localized photonic states. For smaller systems, we do a brute-force open quantum many-body treatment involving the standard quantum master equation approach to a density matrix and a quantum Monte Carlo wave-function method \cite{qjump} for a slightly bigger system. In a semi-classical limit,  we investigate a long 1-D  driven dissipative lattice model. 
Our key findings can be categorized as follows: (i) There exists an optimal window of drive strength where an indefinitely long-lived localized steady state is obtained. As a consequence of this, we get interesting phase diagrams. (ii) In this open system, it is possible to indefinitely sustain a self-trapped/localized state even at a light-matter coupling which is smaller than the critical strength estimated in the fully closed system counterpart. (iii) The achieved self-trapped state is independent of the initial state preparation which is beneficial especially from an experimental perspective. 

\begin{figure}[b]
  \centering
   \includegraphics[width=3.5in]{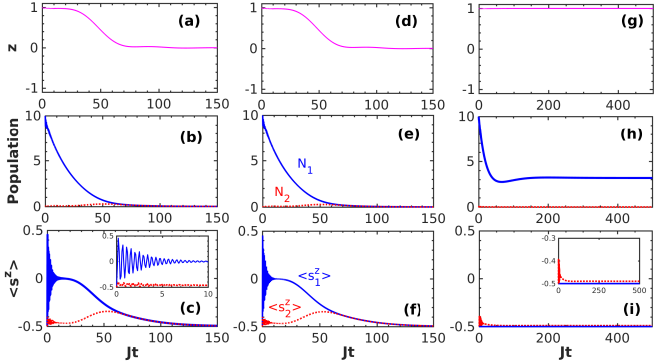}
  \caption{Quantum dynamics for two coupled cavities are presented for initial state $\{{10}, 0, -1/2, -1/2 \}$ and $\kappa=\gamma=.04J$. The value of $g_c$ is $2.8 J\sqrt{10}$.
  (a), (b), and (c) are undriven cases when
  $g_1=g_2=2g_c$. (d), (e), and (f) are drawn for $d_1=.04J, d_2=0$ and $g_1=g_2=2g_c$, i.e., light-matter coupling in both cavities are turned on. (g), (h), and (i) are cases when $g_1=0, g_2=2g_c$ and same driving intensities as in (d). This demonstrates that indefinitely self-trapping can be achieved when one of the cavities does not have light-matter coupling.}
  \label{fig5}
\end{figure}

\textit{Model and Approaches:}\label{model} The Hamiltonian defined for cavity-QED arrays in a rotating frame of drive frequency ($\omega_p$) is given by
\begin{eqnarray}
 H&=& \sum_{i} H_i-J \sum_{\langle ij \rangle} (a_i^{\dagger}a_{j}+h.c)
 \label{ham_coup}
\end{eqnarray}
where
\begin{eqnarray}
 H_i&=& (\omega_0-\omega_p) s^z_i+(\omega_c-\omega_p) a^{\dagger}_ia_i+g_i(a^{\dagger}_i s_i^-+a_is_i^+)\nonumber \\
 &&~~~~~~~~~~~~~~~~~~~~~~~~~~~~~~+d_i(a_i+a^{\dagger}_i).  
 \label{ham_coup_rot}
\end{eqnarray} 
Here, $\omega_0$ and $\omega_c$ are the characteristic frequencies of the qubit (embedded in the 
cavity) and cavity photons, respectively; $a_i$ destroys photon at the i-th cavity and $s^{\alpha}_i$ (where $\alpha\equiv \{x,y,z\}$) is the i-th qubit. $g_i$ is the qubit-photon coupling strength 
that dictates the Rabi oscillations of excitation dynamics and $d_i$ is the drive strength. 
As per the experimental scenario, we introduce the qubit (spontaneous) decay rate $\gamma$ and the photon leakage $\kappa$ of the cavity. Such considerations necessitates us to treat this as an open system whose zero-temperature dynamics could be described by Linblad master 
equation written as
$\dot{\rho}(t)=-i[H,\rho(t)]+\kappa\mathcal{L}[a]+\gamma \mathcal{L}{[\sigma^-}]$
 where  $\mathcal{L}[A]=(2A\rho(t)A^{\dagger}-A^{\dagger}A\rho(t)-\rho(t)A^{\dagger}A$)/2 takes account of the dissipation involved with the qubit or photonic degrees of freedom. $\rho(t)$ is the reduced density matrix of the system. Here we mainly focus on low-temperature behaviour where thermal photon contribution is neglected and only coherent photon drive is employed. Furthermore, we calculate 
 the average photon number and average spin as $N_i={\rm Tr}[a^{\dagger}_ia_i\rho(t)]$ and $\langle s^{\alpha}_i\rangle={\rm Tr}[s^{\alpha}_i\rho(t)]$, 
 respectively. We solve the above Lindblad Master Equation by two approaches (i) The traditional numerical implementation of the above equation and (ii) a quantum Monte-Carlo wavefuntion method which is a powerful technique to deal with a larger Fock space \cite{qjump}. 

It is also imperative to mention an important experimentally feasible limit. 
 In the large photon number limit \cite{tureci_prb} 
one can efficiently approximate the second-order correlation with the 
 product of first-order expectation values. In other words, $\langle a_i s_i^{+}\rangle \approx \langle a_i \rangle \langle s_i^+\rangle$ and so on. Using the above Linblad equation  
 and exploiting the semiclassical approximation, we get 
\begin{eqnarray}
\label{eom_la}
\langle \dot{a_i}(t) \rangle&=&-i(\omega_c-\omega_p)\langle a_i(t)\rangle-ig_i\langle s^-_i(t)\rangle \nonumber \\ &~~+&iJ\big(\langle a_{i+1}(t)\rangle+ \langle a_{i-1}(t)\rangle \big)
\nonumber \\ 
&~~-&\frac{\kappa}{2}\langle a_i(t)\rangle-id_i \\
\label{eom_lb}
\langle \dot{ s^-_i} (t)\rangle&=&-i(\omega_0-\omega_p) \langle s^-_i(t) \rangle+2ig_i\langle a_i(t) \rangle\langle s^z_i(t)\rangle \nonumber \\
&&~~~~~~~~~-\frac{\gamma}{2}\langle s^-_i(t) \rangle \\ 
\langle \dot{s^z_i}(t) \rangle &=&-ig_i(\langle s^+_i(t)\rangle \langle a_i(t) \rangle -\langle a^+_i(t)\rangle \langle s^-_i(t)\rangle)\nonumber \\
&&~~~~~~~~~-\gamma \big(\langle s^z_i(t)\rangle+\frac{1}{2} \big).
\label{eom_l}
\end{eqnarray}
In semiclassical limit the expectation values of the operators can be expressed as complex numbers. To investigate the population trapping phenomenon we rely on the population imbalance, $z(t) = \frac{\sum_{i=1}^{M}  (-1)^{i+1} N_{i}}{\sum_{i=1}^{M} N_i}$ where $N_i = |\langle a_i \rangle|^2$. We designate 
 the state of the system as $\{N_i, \langle s^z_i\rangle \}$.  For $M=2$, we start with the configuration $\{{N}, 0, -1/2, -1/2 \}$ and analyze the value of z with 
 varying g, $d_i$, $\kappa$ and $\gamma$. Using canonical transformation, it has been perturbatively shown that the Jaynes-Cumming interaction induces on-site photon-photon coupling that provides nonlinear contribution 
 to the bare photon dynamics \cite{blaise}. This perturbative regime (dispersive case) is however, not our subject of focus here. 
 For resonant case, the effective anharmonicity is expected to be the strongest. Beyond a critical coupling $g_c\approx 2.8 \sqrt{N}J$, it has been established that a delocalization-localization 
 transition occurs\cite{tureci_prb}. Here we analyse the role of dissipation, drive, and nonlinearity together and present the results in next section.\\
 
 \textit{Results:} In this section, we present results for both (i) the fully quantum case and (ii) the semiclassical limit. Throughout the paper we stick to the near-resonant case $\omega_0-\omega_p=\omega_c-\omega_p=0.01J$.\\

\begin{figure}[t]
  \centering
   \includegraphics[width=1.5in,angle=90,angle=90,angle=90]{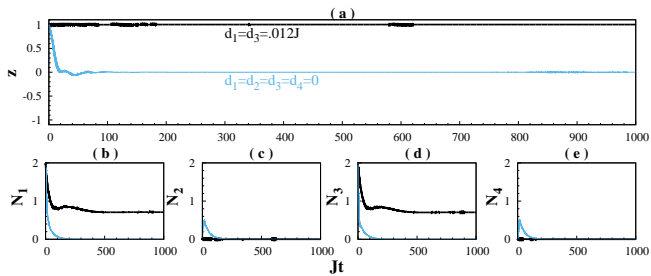}
  \caption{Quantum dynamics for a 1-D chain of four cavities, when 
  the initial conditions are $\{2,0,2,0\}$ for cavity photon population and $\{-1/2,-1/2,-1/2,-1/2\}$ for the qubits. The value of $g_c$ is $2.8 J\sqrt{2}$.
  The results are none-the-less independent of initial conditions as will be shown in Fig.~\ref{fig:emp} (and in supplementary material\cite{supp}).
  (a) Dynamics of $z$ is presented for $\gamma=\kappa=.02J$, $M=4$, $g_1=g_3\approx4.5g_c$, and $g_2=g_4=0$. The undriven case having $d_i$ values $0$ (thin light blue) is contrasted with a case 
  where $d_1=d_3=.012J$ and $d_2=d_4=0$ (thick black). (b), (c), (d), (e) describe the photon population of four cavities, respectively.  
 }
  \label{fig8}
\end{figure}

\begin{figure}[t]
  \centering
   \includegraphics[width=1.2in,angle=90,angle=90,angle=90]{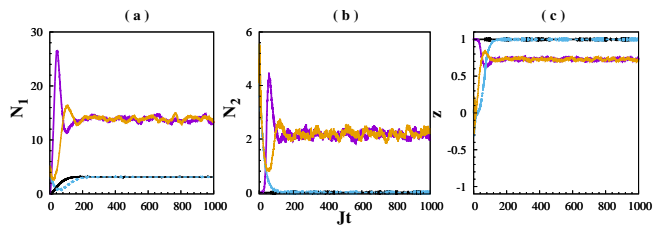}
  \caption{Quantum dynamics and demonstration of initial-condition independence. Here $M=2$, $\gamma=\kappa=.04J$, and $g_1=0, g_2\approx 1.2 g_c$ (where $g_c$ is same as Fig.~\ref{fig5}, i.e., $g_c=2.8 J\sqrt{10}$).
  Plots are shown for two distinct initial conditions: (i) $N_1=N_2 =0$ (both cavities are initially vacuum presented by black and purple) and 
  (ii) $N_1=N_2 =5$ (presented by dashed blue and yellow).
  For each initial condition we show results for two $\{d_1,d_2\}$ compositions : (i) $d_1=.04J, d_2=0$ (black and dashed blue) and (ii) $d_1=.2J, d_2=0$ (purple and yellow). 
   (a), (b), and (c) plot $N_1$, $N_2$, and $z$ dynamics, respectively.
  The initial condition independence can be seen in our results.  The initial conditions for qubits in all cases were $\{-1/2,-1/2\}$ but we find that the initial condition independence also holds for different initial conditions of the qubits (see supplementary material \cite{supp})}
  \label{fig:emp}
\end{figure}

\textit{(i) Results from Quantum Simulation: }\label{quantum}Here we carry on full quantum treatment for a case with relatively small number of photons and solve the Lindblad Master equation in a many-body framework. The light-matter interaction $g_i$ will be written in terms of the critical value \cite{tureci_prb} of the corresponding unitary case ($g_c \sim 2.8 J \sqrt{N}$). This convention is used as a reference point and our work focuses on the non-unitary case. The undriven case in Fig. \ref{fig5} (a), (b), and (c) shows that dissipation limits the lifetime of localization (as will be seen even in the semiclassical picture in a subsequent discussion). Comparing Figs. \ref{fig5} (b) and (c) we see that the qubit energy in the right cavity increases whenever the photon population $N_2$ shows a bump in 
Fig. \ref{fig5} (b). 
The inset of Fig. \ref{fig5} (c) also addresses rapid decay of Rabi oscillations. In Figs. \ref{fig5} (d), (e), 
and (f) we drive the left cavity coherently and no enhancement of localization is observed. In fact, we find (not shown here) only a small enhancement, even at much larger driving intensities ($g_1,g_2 \neq 0$). 
This makes the quantum scenario strikingly different from the semiclassical case where quantum correlation is 
ignored. Such a purely quantum effect pushes away external photons making the drive ineffective. 
The dynamics of such photon-photon correlation is presented in the supplementary material\cite{supp} (Fig. S1).
Now, we nullify this correlation by switching off the cavity-qubit 
coupling (i.e., $g_1=0$) only for the left cavity and plot the results in Fig. \ref{fig5} (g), (h), and (i). Fig. \ref{fig5} (g) shows perfectly stabilized localization 
at the same $d_1$ value used in Fig. \ref{fig5} (d); Fig. \ref{fig5} (h) supports the attainment of nonzero population for the left-cavity steady state and Fig. \ref{fig5} (i) 
shows that the isolated left qubit remains in its ground state and the right qubit shows some dynamics because $g_2 \neq 0$ [see inset of Fig. \ref{fig5}(i)]. Therefore, we have demonstrated that by carefully engineering the Hamiltonian, one can achieve an indefinitely long-lived and perfectly self-trapped state in a fully quantum system.

We scale up the quantum case to a 1-D chain of four cavity-QEDs and plot the results in Fig. \ref{fig8} when only the first and third cavities are 
initially populated. 
Even for $g_i$ values deep into the localized phase, $z$ shows long-time oscillations when photon population is below a critical value \cite{tureci_prb}. 
With finite $\gamma$ and $\kappa$ such oscillations of $z$ decays to zero in Fig. \ref{fig8} (a) (light blue). On the other hand, the driven-dissipative 
situation (thick black) localizes the photon populations in first and third cavities [see Figs. \ref{fig8} (b), (c), (d), and (e).].
We chose $M=4$ and tested for convergence issues. This figure demonstrates the indefinitely long-lived localization in an extended open quantum system.  
We would like to point out a remarkable observation here. In the unitary case (neither drive nor dissipation), one would expect no self-trapping (i.e., localization) for 
small photon number\cite{tureci_prb}. However, the open analog of the same setup can be made to perfectly localize in the steady state by a unique balance between drive and dissipation. 
Fig.~\ref{fig:emp} demonstrates initial-condition independence in this open quantum system. In addition to this being interesting, it is also more experimentally practical to start with a vacuum 
state in the cavities and $s^z_i =-1/2$ for the qubits.  Such a setup when driven out of equilibrium leads to interesting non-equilibrium steady states. In Figs. \ref{fig:emp} (a), (b), and (c) 
we find that an optimum drive $d_1=0.04J$ produces a self-trapped steady state, whereas an overdrive $d_1=0.2J$ populates the undriven second cavity destroying strong self-trapping [see Fig. \ref{fig:emp}(c)] leading to only partially self-trapped state.

\begin{figure}[b]
  \centering
   \includegraphics[width=3.5in]{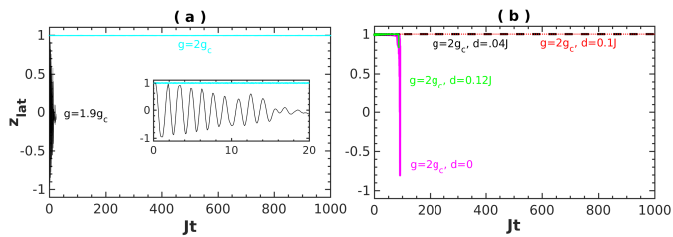}
  \caption{Semiclassical dynamics for population imbalance in a 1-D lattice consisting 100 identical cavity-QEDs (each hosting a qubit) is presented with $g_i$ and $d_i$ values mentioned with similar colors as the respective lines.
  In all the cases every odd cavity is initialized with 20 photons and even ones are kept vacant.
  $\kappa$, $\gamma$, and $g_i=g$ values are kept identical for each cavity in the array.
  (a) Depiction of delocalization-localization transition for a closed-system case (no drive and dissipation). (b) Open system case (i.e., with drive and dissipation) with $\kappa=\gamma=0.04J$ which reflects stabilization (indefinitely long lived) of self-trapped state 
  with appropriate drive. Here, $g_c = 2.8 J \sqrt{20}$.}
  \label{fig4}
\end{figure}

\begin{figure}[b]
  \centering
 \includegraphics[width=3.5in]{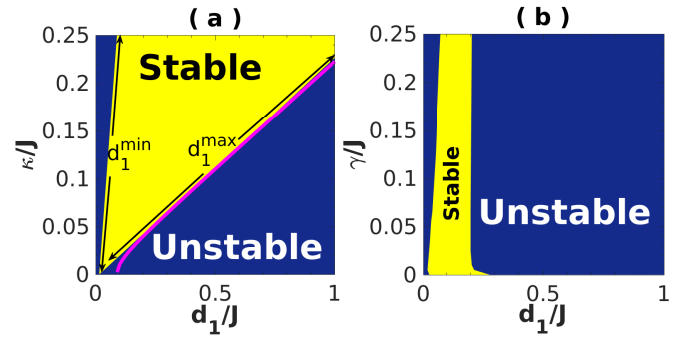}
  \caption{Semiclassically obtained  phase space description of long-lived trapped state. 
  (a) Description in $d_1-\kappa$ plane when $\gamma=.04J.$
  (b) Description in $d_1-\gamma$ plane when $\kappa=.04J$. 
  In both the figures, cavity system is initialized to a state $\{{20},0,-1/2,-1/2\}$ and 
  fixed coupling for both the cases $g_1=g_2=2g_c$. Here yellow region depicts (marked as `stable') the phase where localization persists in the steady state, whereas, the blue region (marked as `unstable') represents delocalization. 
  The limiting values $d^{min}_1$ and $d^{max}_1$ lie on the upper and lower boundaries of `stable' region.  In (a), the magenta line gives an analytical prediction (Eq.~\ref{estimate}) for the curve $\kappa(d_1)$ representing the upper boundary between self-trapped and non-self trapped phases. }
  \label{fig3}
\end{figure}
\begin{figure}[b]
  \centering
   \includegraphics[width=3.5in]{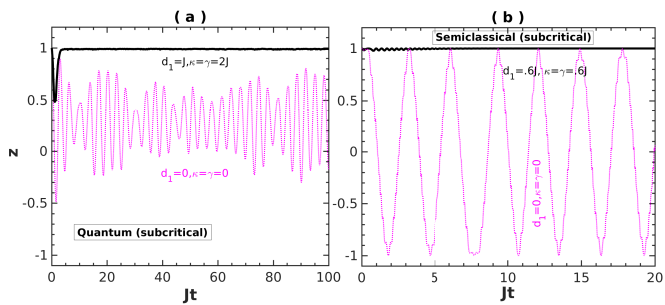}
  \caption{Demonstration of indefinitely long lived self-trapping even at sub-critical interaction in both the quantum and the semiclassical treatments. Attaining sustained self-trapping for subcritical light-matter interaction when a 2-cavity system is initialized at $\{20,0,0.5,0.5\}$. a) Quantum case where $g_1=0$, $g_2=0.4g_c$. b) Semiclassical case where $g_1=g_2=0.9g_c$. 
 Here, $g_c$ is given by the fully closed system counterpart, i.e., $g_c = 2.8 \sqrt{20}J$. Drive and dissipation rates are mentioned with respective colors.}
  \label{figsub}
\end{figure}

\textit{(ii) Semiclassical Results: }\label{semiclassic} In a semiclassical framework we neglect the quantum correlations and make use of Eqs.~\ref{eom_la}, \ref{eom_lb}, \ref{eom_l}. We present results for a 1-D lattice of identical cavities. We investigate $M=100$ cavities each hosting a qubit with identical couplings $g_i=g$.
Although we can tackle much larger lattices, the essential physics remains same.
Without loss of generality, we choose initial conditions such that 
all the odd cavities  are populated and all even cavities are vacant. The defined population imbalance $z$
is plotted in Figs. \ref{fig4} (a) and (b). The closed system case is shown in Fig. \ref{fig4} (a) 
which shows localization when $g \approx 2 g_c$ with $g_c=2.8\sqrt{N}J$ (where $N$ is the initial photon number on every odd site). In contrast to the two-cavity case ($M=2$) in Supplementary Fig. S2, here we need $g=2g_c$ for 
localization (the factor of $2$ is essentially the coordination number). In Fig. \ref{fig4} (b) the driven-dissipative scenario is presented where 
appropriately driven odd sites stabilize the localized state. These results indicate that drive and dissipation can be carefully designed to create localized steady states in large-scale systems. 
At $t_{break}$ (where the self trapping is just disrupted) rapid oscillation of $z$ sets in and this indicates photon tunneling throughout the lattice. We show only few oscillations after $t_{break}$ 
that converged for various numerical precisions. As we advance in time, subsequent oscillations become too sensitive to precision and barely give any relevant physical insight. This regime is anyway not of our interest as the self-trapped phase is already destroyed. Similar and detailed analysis for the case of $M=2$ is presented in supplementary material\cite{supp} (Fig.~S2).
Through our semiclassical analysis, we also find that  as long as the system is in a self-trapped phase, the results are insensitive to the details of the initial state, thereby making it more experimentally feasible.
The conditions for the semi-classical approximation to hold is in tune with the current experimental setups \cite{raftery}. From the above diagram (Fig.~\ref{fig4}), one can notice that an intricate interplay between drive, dissipation and interaction can lead to the existence of an optimal window of drive strengths where localized steady states can exist. This naturally leads to an interesting question regarding the existence of a phase diagram in such systems. We investigate this for the case of $M=2$ without loss of generality.

Fig.~\ref{fig3} ($M=2$) demonstrates an interesting phase diagram. In Fig.~\ref{fig3}a we see that there is a minimum drive $d_1^{min}$ and a maximum drive $d_1^{max}$ 
for a given value of cavity decay $\kappa$ (keeping all other parameters fixed) and they define the limiting drives for indefinitely long-lived localization. The minimum drive can be understood as the least amount of driving needed to assure reasonable population in the first cavity in comparison with the second one. The maximum drive suggests that the driving beyond a point overcomes the blockade due to interaction $g_2$ which 
leads to population increase in the other cavity thereby destroying self-trapping. This upper limit of drive can be analytically discussed. Assuming that there is perfect 
self-trapping  at steady state, one arrives at the below analytical equation for $\kappa(d_1)$ which is given by the magenta line in Fig.~\ref{fig3} (a),
\begin{equation}
\kappa(d_1) =  2 \sqrt{2.8^2\Big ( \frac{d_1}{g}\Big)^2 -(\omega_0-\omega_p)^2},
\label{estimate}
\end{equation}
where $g_1=g_2=g$.
 Eq.~\ref{estimate} is derived from Eqs.~\ref{eom_la} \ref{eom_lb}, \ref{eom_l} with the assumption that the cavity 2 is effectively disconnected since $\langle a_2 \rangle  \sim 0$. Moreover, we assume that the qubit in the first cavity plays no role as   $\langle s_1^{-} \rangle  \sim 0$. This is of course an approximation but fits remarkably well with the numerical results. Indeed, in the regime where the analytical curve agrees with the numerical phase boundary, the numerical results show that  $\langle s_1^{-} \rangle  $ and $\langle a_2 \rangle$ are negligible.
$d^{max}_1$ is determined by the interplay of $g$ and $N_1$.  Given a particular $g$, as long as $d_1$ does not produce $N_1$ that renders $g$ subcritical, (i.e., $g < 2.8\sqrt{N_1}J$) self-trapped steady state exists. For further analysis on the two-cavity case see Figs. S2 and S3 in the supplementary material \cite{supp}. We would like to point out that we see minor discrepancy between analytical curve and the phase boundary from numerics (Fig. \ref{fig3} a) in the regime where $\kappa$ is small. The case with fixed $\kappa$ and varying $\gamma$ is depicted in Fig. \ref{fig3} (b), where $d^{max}_1$ remains almost fixed with increasing $\gamma$. This is a mere consequence of negligible photon loss via spontaneous decay channel ($\gamma$) of the qubit (at a fixed $\kappa$).

 One remarkable difference between the quantum and semiclassical treatments is the effectivity of $d_1$. The strong antibunching of photons in Fig. \ref{fig5}(d) (with $g_1\neq0$) 
makes the cavity resistant to external drive. This effect is absent in semiclassics where we neglect second order correlation between the qubit and photons at large $N$ limit. An important finding of our work is that both in the quantum and semiclassical treatments, indefinitely long-lived self-trapping can be attained even at subcritical interaction strength (where by critical we mean, the value estimated from the closed system counterpart) if we have optimally engineered driven-dissipative systems (see Fig.~\ref{figsub}).

\textit{Discussion and conclusion:}\label{conclusion}
In this manuscript, we have studied an open system consisting of 1D lattice of cavities each hosting a qubit. This system is further subject to drive and dissipation.  By careful interplay between dissipation, drive, and cavity-qubit coupling, we have predicted a parameter regime where an indefinitely long-lived self-trapped state exists. First, we solve the fully quantum problem for the case of two cavities (driven-dissipative Jaynes-Cummings dimer). We then show exact quantum results for the case of a chain of 4 cavities. In this case also, we see indefinitely long lived localized states. 
For smaller quantum systems, we used quantum master equation solutions for density matrix, whereas, slightly larger systems are dealt with Monte-Carlo wavefunction techniques.
Next, in order to deal with a very large-scale system, we employ mean-field treatments (semiclassical) which becomes accurate for 
large-cavity arrays with large number of photons. Here too we get similar physics as in the quantum case. 
Our work is relevant to achieve indefinitely long-lived localized states by striking a delicate balance between drive, dissipation and interactions. 
Such exotic steady states are independent of initial system preparation, which is advantageous for experiments. Furthermore, remarkably an optimum drive window for sustained self-trapping 
is available even if the light-matter interaction is tuned at subcritical values. Our analysis of engineered driven-dissipative cavity system  
will help in precisely accessing localization/delocalization phases and will also be paramount for well-controlled photon transport in a cavity array. \\
\linebreak 
Future outlook involves developing quantum methods for dealing with even larger systems. This could be key for investigating indefinitely long-lived many-body localization in driven-dissipative systems. The role of counter-rotating terms (for e.g., the Rabi-Hubbard) in the physics of self-trapping still remains unexplored. Needless to mention, the physics of localization remains fascinating in two or more dimensions and artificially engineered systems with deformable lattices \cite{houck_nature}. 

\textit{Acknowledgements:}
We would like to thank J. Keeling for useful discussions. M. K. gratefully acknowledges the Ramanujan Fellowship SB/S2/RJN-114/2016 from the Science and Engineering Research Board (SERB), Department of Science and Technology, Government of India. M. K. also acknowledges support the Early Career Research Award, ECR/2018/002085  from the Science and Engineering Research Board (SERB), Department of Science and Technology, Government of India. MK would like to acknowledge support from the project 6004-1 of the Indo-French Centre for the Promotion of Advanced Research (IFCPAR).

%% file: self_trapped_suppl.tex
\title{ {\bf Supplemental Material for}\\``Engineering indefinitely long-lived localization in coupled cavity-QED arrays''}
\author{Amit Dey}
\author{Manas Kulkarni}
\address{International Centre for Theoretical Sciences, Tata Institute of Fundamental Research, Bengaluru -- 560089, India}
\maketitle

\section{Bunching and Antibunching effect of the driven cavity}
\begin{figure}[htbp] 
   \centering
   \includegraphics[width=3.5in]{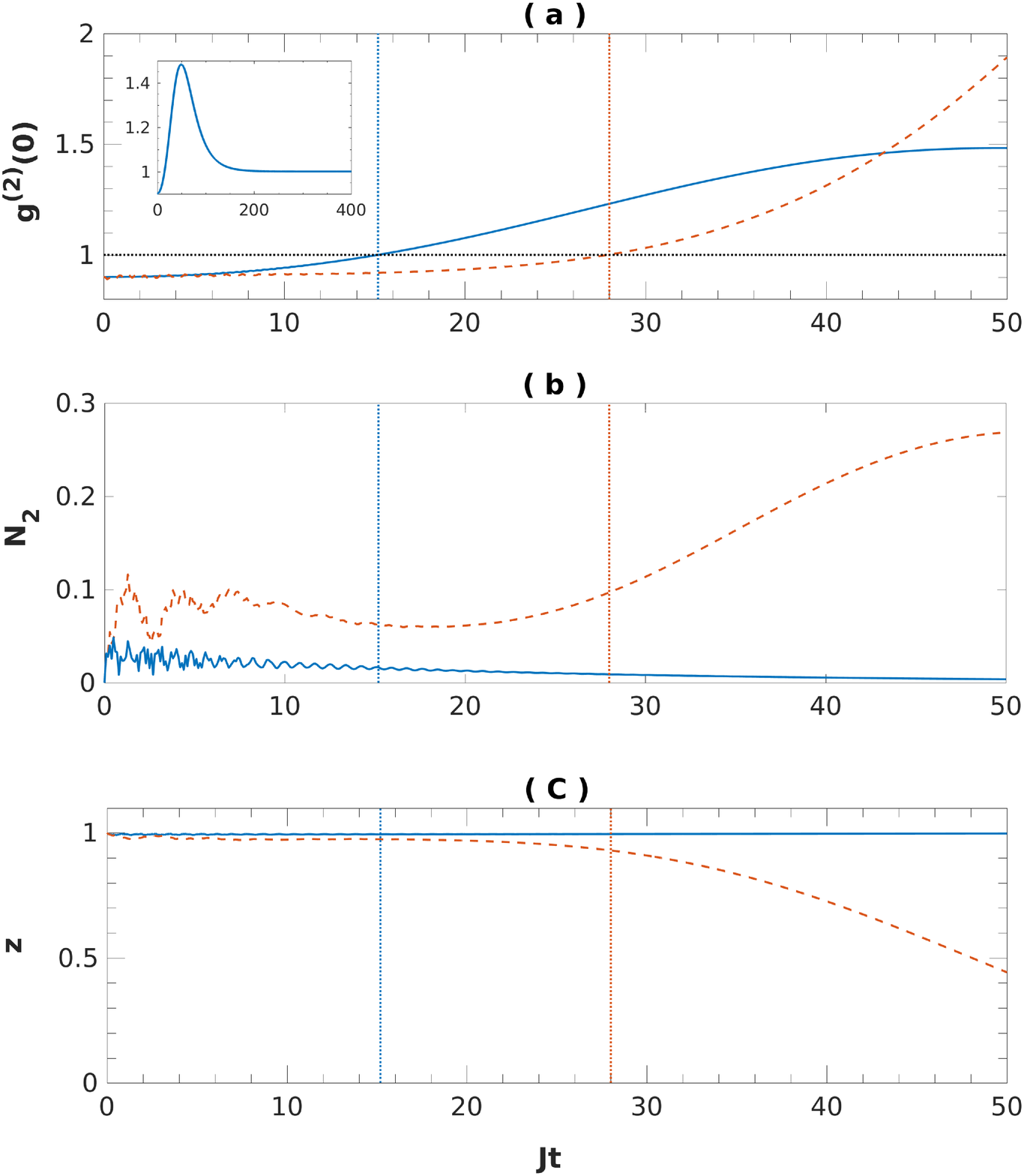} 
   \caption{(a) Dynamics of quantum correlation function $g^{(2)}(0)$. Horizontal dotted black line in (a) marks $g^{(2)}(0)=1$ value.
                 (b) $N_2$ dynamics. 
  (c) Short time dynamics of z is shown for the two cases. 
  Vertical dotted lines of respective colors mark the times when $g^{(2)}(0)$ attains 1 in all figures. Parameter values are same as in Fig. 2 (d) (orange dashed) and Fig. 2 (g) (solid blue) of main text, respectively. }
   \label{fig7}
\end{figure}

To sharpen our explanation for effectivity of drive in Fig.~2 (of the main manuscript), we plot temporal behavior 
of undelayed second order correlation, given by $g^{(2)}(\tau){\Large |}_{\tau=0}=\langle a^{\dagger}_1(t+\tau) a^{\dagger}_1(t) a_1(t+\tau) a_1(t) \rangle / \langle a^{\dagger}_1(t)a_1(t) \rangle^2{\Large |}_{\tau=0}$ (left cavity). Fig. \ref{fig7} (a) shows that, antibunching to bunching 
transition occurs earlier when the cavity-qubit coupling is switched off for the left cavity. On the other hand, when light-matter interaction $g\neq0$ for both the cavities, this transition appears at time where 
the blockade in the right cavity is already broken and the localization starts getting spoiled [see Figs. \ref{fig7} (b) and (c)]. Furthermore, the inset describes coherent light at long times for the 
former case; this is obvious as the steady state describes empty right cavity and a left cavity with only noninteracting photons. We thus demonstrate that antinbunching in the first cavity (which is also the driven) is enhanced when $g_1 \neq 0$ which thereby disallows self trapping. In other words, the cavity that is driven should have bunching effect in order to support sustained self-trapping. 
\section{Semiclassical results for two coupled cavities }
\begin{figure}[h]
  \centering
   \includegraphics[width=3.5in]{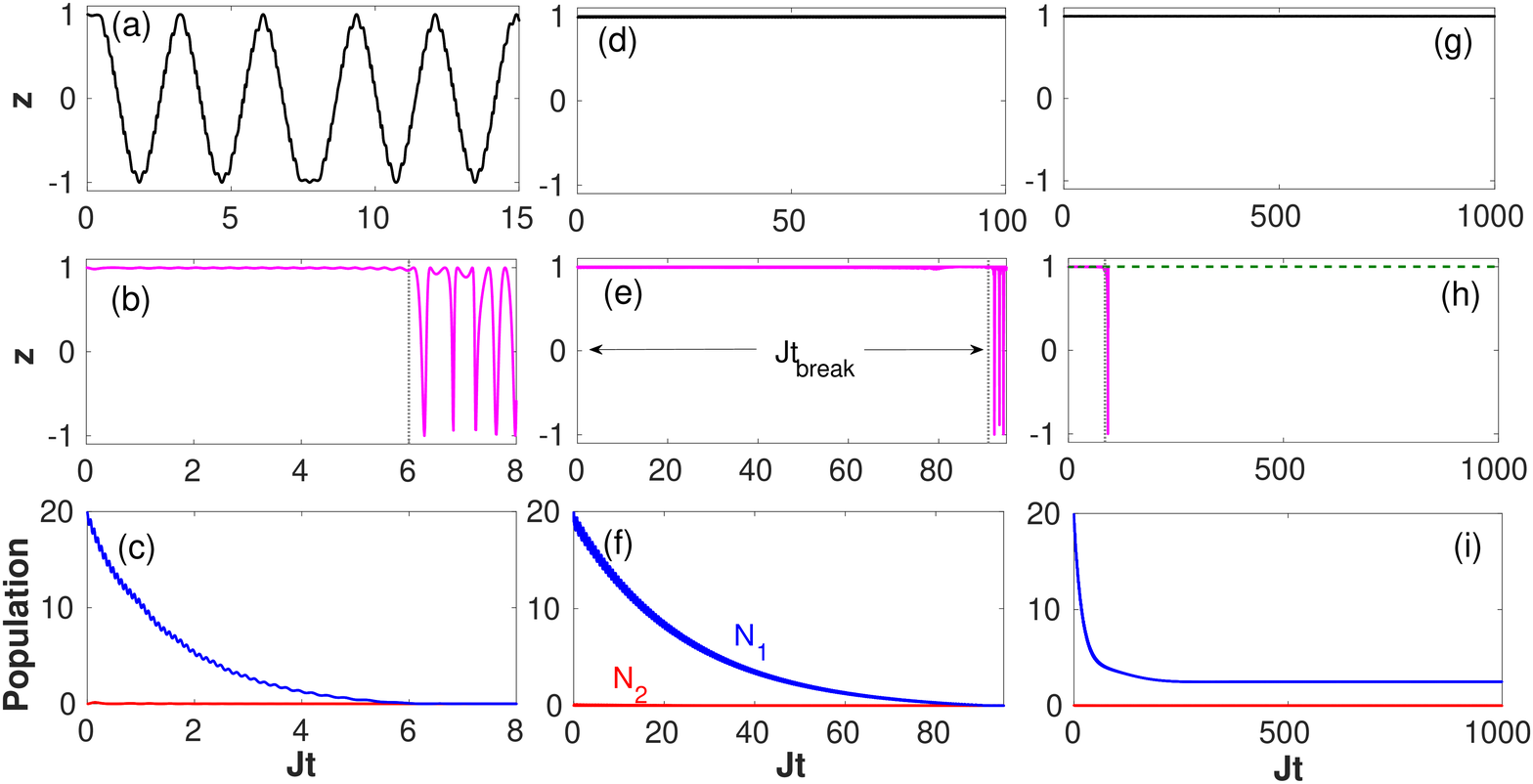}
  \caption{Semiclassical behavior of population dynamics of two coupled cavities with varying $\gamma$, $\kappa$, $g_1=g_2=g$, and drive. We choose $g_c = 2.8 J \sqrt{20}$.
  (a) $g=0.9g_c$, $\gamma=0$, $\kappa=0$, and $d_1=d_2=0$  (closed system with subcritical light-matter coupling). 
  (b) $g=0.9g_c$, $\gamma=0.6J$, $\kappa=0.6J$, and $d_1=d_2=0$ (dissipative system with no drive). 
  (d)  $g=g_c$, $\gamma=0$, $\kappa=0$, and $d_1=d_2=0$ (closed system with critical light-matter coupling). 
  (e)  $g=g_c$, $\gamma=0.04J$, $\kappa=0.04J$, and $d_1=d_2=0$ (dissipative system with no drive, and with critical light-matter coupling). 
  (g) $g=2g_c$, $\gamma=0$, $\kappa=0$, and $d_1=d_2=0$ (closed system with light-matter coupling greater than $g_c$).
   (h) $g=2g_c$, $\gamma=0.04J$, $\kappa=0.04J$. $d_1=d_2=0$ (solid magenta) and $d_1=0.04J, d_2=0$ (dashed green). (c), (f), and (i) are the population dynamics 
  for left and right cavities for parameter specifications as in (b), (e), and (h), respectively. The dotted grey vertical lines in (b), (e), and (h) mark the break time $t_{break}$ when the localization  
  starts getting destroyed.}
  \label{fig1}
\end{figure}

 In Fig. \ref{fig1} we plot dynamics of z for different couplings g, $\gamma$, $\kappa$, $d_l$, and $d_r$, when only the left cavity is loaded with photonic population with 
 initial state $\{20,0,-1/2,-1/2\}$. 
 Fig. \ref{fig1} (a) depicts the lossless coherent oscillation of 
 undriven photonic population for $g=0.9g_c$; whereas, in Fig. \ref{fig1} (b), introduction of strong dissipation localizes z for a limited time and the population decays thereafter. The moment the localized state 
 gets destroyed (marked by $Jt_{break}$), we get rapid oscillations due to decayed population. Fig. \ref{fig1} (c) shows the dynamics of population distribution in the respective cavities and fast oscillations beyond the break time. 
 However, as dissipation reduces N, the modified critical condition suits the subcritical g we start with and localization is established.  Similar message is reflected in Figs. \ref{fig1} (d), (e), 
 and (f) but for a case with critical coupling $g=g_c$. Comparing Figs. \ref{fig1} (d) and (e), we see that the dissipative setup destabilizes the already localized regime by photon leakage. The case deep within the localized 
 regime is presented in Figs. \ref{fig1} (g), (h), and (i). In Fig. \ref{fig1} (h) we notice that, when left cavity is coherently driven, the localized state persists even at long times. Furthermore, 
 Fig. \ref{fig1} (i) demonstrates the balancing of drive and dissipation and attainment of a steady state with finite $N_1$ and $N_2\approx0$.
\begin{figure}[b] 
   \centering
   \includegraphics[width=3.5in]{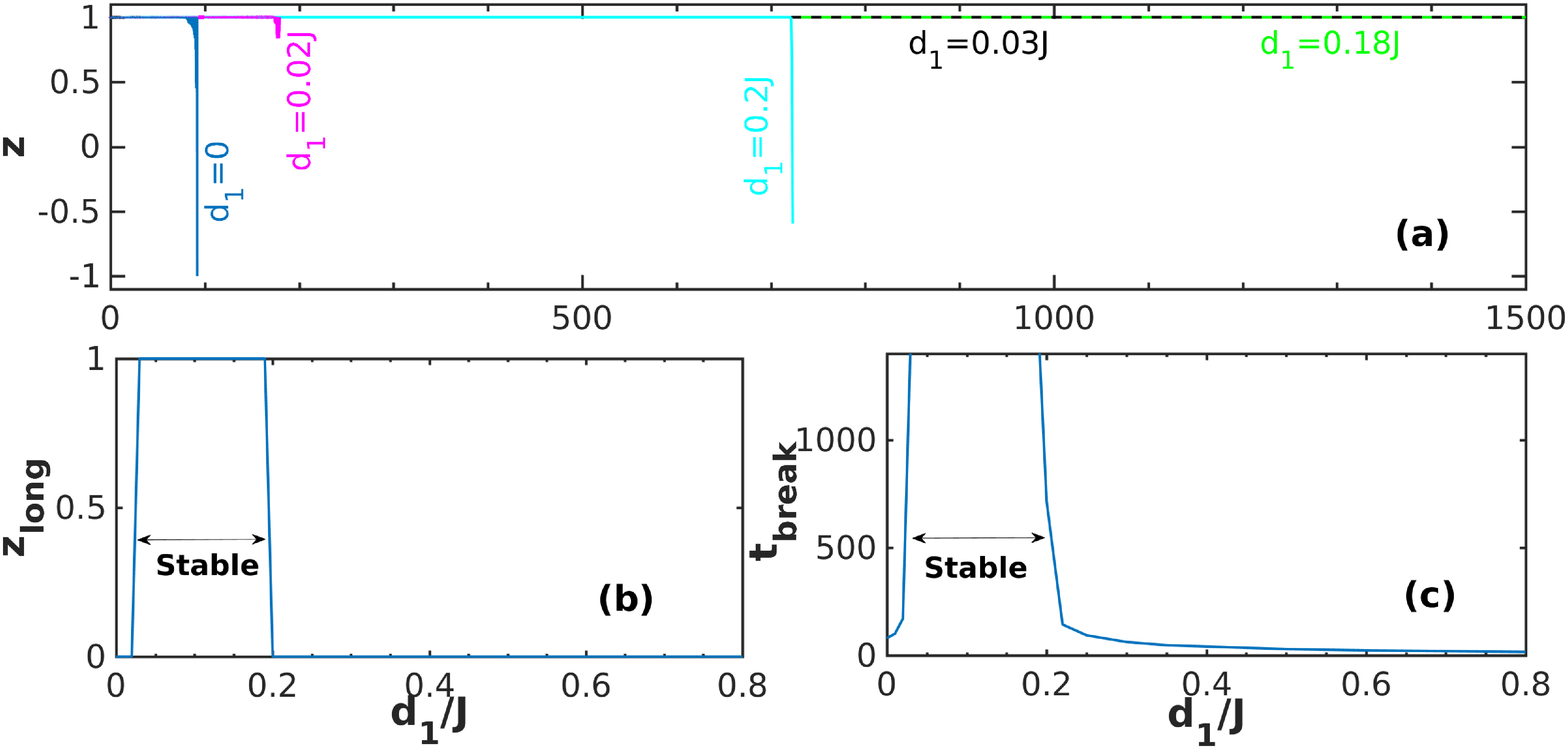} 
   \caption{(Semiclassical behaviour) (a) Temporal behavior of z for various $d_1$ values when $g=2g_c$ and $\gamma=\kappa=0.04J$. (b) Long-time values of z =${\rm z_{long}}$ and (c) $t_{break}$ for varying $d_1$ are plotted. Here $g_c = 2.8J\sqrt{20}$. These results (b) and (c) show that semiclassical method predicts an abrupt change between no self-trapping and $100\%$ trapped state.}
   \label{fig2}
\end{figure}

\section{Optimum drive for stable localization}
Fig. \ref{fig2} depicts the temporal stability of localization with varying $d_1$ when the system resides deep into the self-trapped regime. It is evident from Figs. \ref{fig2} (a) and (b) that there exists an optimum range 
of $d_1$ that reinforces $N_1$ as to stabilize z over long times. $Jt_{break}$ is plotted in Fig. \ref{fig2} (c) where the upper limit of y axis can be taken much greater than 1400 as in the stable region 
 $t_{break}$ actually diverges; we set this limit just for representation purpose.

Here we have constructive and destructive roles of photon drive at the two boundaries of optimum range, respectively. 
 { At lower bound drive wins over dissipation, whereas it spoils critical $g$ at the upper bound. Although the two-cavity case is a special case of the 1-D lattice (Fig. 5) in the main text, they reflect similar 
 physics.}
\section{Initial-state independence of steady state}
\begin{figure}[t] 
   \centering
   \includegraphics[width=1.3in,angle=90,angle=90,angle=90]{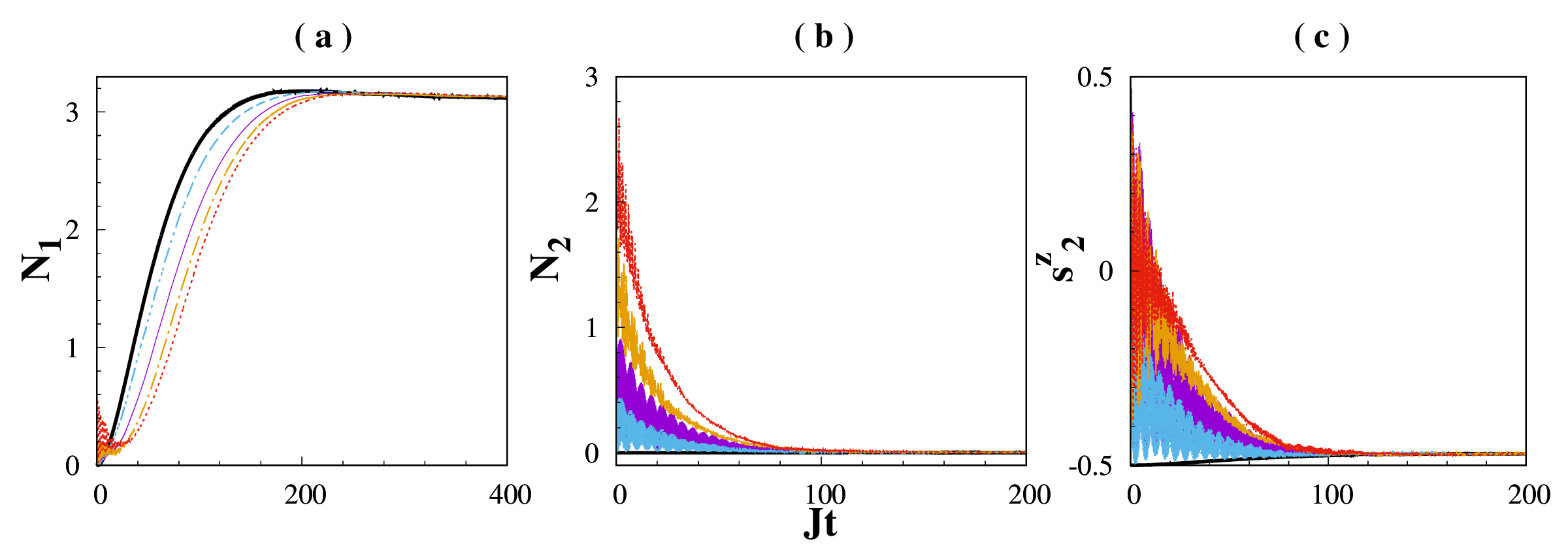} 
   \caption{(Exact quantum simulation) Photon and qubit dynamics for various initial photon number distribution and qubit states. Here $\kappa=\gamma=.04J$,
   $d_1=.04J, d_2=0$, and $g_1=0, g_2=1.2*2.8\sqrt{10}J$. Various preparations are as follows. 
   (i) $\{0,0,-1/2,-1/2\}$ (thick solid black), (ii)$\{0,0,-1/2,1/2\}$ (thin solid purple), (iii) $\{0,0,-1/2, 0\}$, i.e., the qubit state in cavity 2 is equal superposition of excited and ground state (dashed blue), 
   (iv) $\{0,2,-1/2,-1/2\}$ (dashed dotted yellow), and (v) $\{0,2,-1/2,1/2\}$ (dotted red). At long time the uniqueness of steady state is evident.}
   \label{fig3_sup}
\end{figure}

Here we present full quantum result showing that the steady state values are independent of initial photon number distribution and qubit states. In Fig. \ref{fig3_sup} (a), (b), and (c) various preparations approach 
unique steady state at long times. The final state is only dictated by $\kappa$, $\gamma$, $d_1$, and the light-matter coupling.
